\documentclass[10pt,letterpaper]{article}
\pdfoutput=1

\usepackage{graphicx}

\renewcommand{\baselinestretch}{2}

\usepackage[margin=2.5cm]{geometry}
\usepackage[small,font=sf]{caption}

\begin{document}

\setlength{\parindent}{0em}

{\sf \LARGE Tunable few-electron double quantum dots and Klein tunnelling in
ultra-clean carbon nanotubes}

\vspace{0.5em}

G. A. Steele$^1$, G. Gotz$^1$, \& L. P. Kouwenhoven$^1$

\vspace{1em}

{\em $^1$Kavli Institute of NanoScience, Delft University of
Technology, PO Box 5046, 2600 GA, Delft, The Netherlands.}

\vspace{1em}

{\bf Quantum dots defined in carbon nanotubes are a platform for both
basic scientific
studies\cite{Hanson2007Spins,Mason2004Local,Sapmaz2006Excited,Rgensen2006Single,Graber2006Molecular}
and research into new device applications\cite{Loss1998Quantum}. In
particular, they have unique properties that make them attractive for
studying the coherent properties of single electron
spins\cite{Petta2005Coherent,Koppens2007Universal,KuemmethCoupling,Nowack2007Coherent,Bulaev2008Spinorbit}. To
perform such experiments it is necessary to confine a single electron
in a quantum dot with highly tunable barriers\cite{Hanson2007Spins},
but disorder has until now prevented tunable nanotube-based
quantum-dot devices from reaching the single-electron
regime\cite{Mason2004Local,Sapmaz2006Excited,Rgensen2006Single,Graber2006Molecular}. Here,
we use local gate voltages applied to an ultra-clean suspended nanotube
to confine a single electron in both a single quantum dot and, for the
first time, in a tunable double quantum dot. This tunability is
limited by a novel type of tunnelling that is analogous to that in the
Klein paradox of relativistic quantum mechanics.}

\setlength{\parindent}{2em}

Single spins in carbon nanotube quantum dots are expected to be very
stable against both relaxation and
decoherence\cite{Bulaev2008Spinorbit}. Nuclear spins, the principal
source of spin decoherence in
GaAs\cite{Petta2005Coherent,Koppens2007Universal}, can be completely
eliminated and, furthermore, a strong spin-orbit interaction recently
discovered in carbon nanotubes\cite{KuemmethCoupling} enables
all-electrical spin
manipulation\cite{Nowack2007Coherent,Bulaev2008Spinorbit}, while
preserving long spin relaxation and decoherence
times\cite{Bulaev2008Spinorbit}. Electron spins in carbon nanotube
quantum dots are therefore attractive for implementation of a quantum
bit (qubit) based on spin for applications in quantum-information
processing\cite{Loss1998Quantum}. In double quantum dot systems,
precise control of the tunnel coupling between the two quantum dots,
and between the quantum dots and the leads attached to them, is
critically important for spin readout
schemes\cite{Fujisawa2002Allowed,Elzerman2004Singleshot,Hanson2007Spins},
and also to prevent loss of spin and phase information through
exchange of an electron with the leads.

Double quantum dots can also be used to explore novel quantum
tunnelling phenomena. In Klein
tunnelling\cite{Klein1929Die,Katsnelson2006Chiral,Trauzettel2007Spin},
for example, an electron tunnels with a high probability through a
long and tall potential energy barrier when the height of the barrier
is made comparable to twice the rest mass of the electron. It is not
feasible to create such a barrier for free electrons due to the
enormous electric fields required, but the low effective rest mass of
the electrons in small bandgap nanotubes makes the observation of such
Klein tunnelling in nanotube devices possible\cite{Trauzettel2007Spin}.

By depositing metallic gates isolated by a dielectric layer on top of
a nanotube, several groups have demonstrated tunable double quantum
dots in nanotubes lying on a
substrate\cite{Mason2004Local,Sapmaz2006Excited,Rgensen2006Single,Graber2006Molecular}.
A disadvantage of this technique is that nanotubes in these devices
suffer from significant disorder induced by the substrate and by the
chemical processing required to fabricate the device. As the electron
density is reduced, this random potential dominates and breaks the
segment of nanotube into multiple disorder-induced ``intrinsic''
quantum dots before reaching the few-electron regime.

Wet etching of the device after fabrication to remove the
substrate-induced disorder has been used previously to obtain single
electron quantum dots in carbon
nanotubes\cite{JarilloHerrero2004Electronhole,Minot2004Determination},
although experience has shown that the yield of such devices is quite
low. Recently, a new fabrication method has been developed for
producing ultra-clean quantum dots in suspended carbon nanotubes with
a high yield in which all chemical processing is done before nanotube
growth\cite{Cao2005Electron}. Studying single quantum dots in these
devices has uncovered new carbon nanotube physics, including a strong
spin orbit interaction due to the nanotube
curvature\cite{KuemmethCoupling} and evidence of Wigner
crystallization of electrons at low
density\cite{Deshpande2008Onedimensional}.  While devices fabricated
in this way are extremely clean, they have some significant
limitations: in particular, the confinement is produced only by
Schottky barriers, which cannot be easily tuned in-situ. Furthermore,
due to an insufficient number of local gates, it has not been possible
to create a tunable double quantum dot in these ultra clean devices.

In order to overcome these limitations, we have developed a new method
of integrating multiple local gates with the ultra clean
fabrication. A schematic of the device is shown in Figure 1. As
described in the Methods section, we grow a carbon nanotube over gates
that are patterned in a thin doped silicon layer. Our current design
provides three independent gates, although fabrication can easily be
modified to include a scalable number of gates inside the trench (see
Supplementary Information). In this letter, we use these three gates in
two different ways. In device D1, with $L$ = 1.5 $\mu$m, the gates are
used to define a single electron and single hole quantum dot where
electrons and holes are confined by tunable {\em pn}-junctions instead
of Schottky contacts. In device D2, with $L$ = 300 nm, we rely on
tunnel barriers from the Schottky contacts, but now use the three
gates to create a tunable single electron and single hole double
quantum dot.

In all previous measurements of quantum dots in carbon nanotubes
containing a single electron, carriers were confined by Schottky
barriers formed at the metal
contacts\cite{KuemmethCoupling,JarilloHerrero2004Electronhole}, or by
potentials defined from trapped oxide
charges\cite{Minot2004Determination}. In figure 2, we demonstrate a
single electron quantum dot defined only by gate voltages. We begin by
applying a negative voltage to the splitgates, creating a p-type
nanotube source and drain on top of the oxide. Sweeping the backgate
voltage V$_\mathrm{BG}$, shown in figure 2a, the current initially
shows weak modulations from resonances in the leads when the suspended
segment is p-type ($ppp$ configuration), and is then completely
suppressed as the suspended segment is depleted ($pip$
configuration). As we sweep further, we form a {\em pnp} quantum dot
showing clean Coulomb blockade, where single electrons in the
suspended segment are confined by {\em pn} junctions to the
leads. Figure 2c shows a stability diagram as a function of both
backgate and bias voltage, demonstrating that we have reached the
single electron regime. As the confinement potential and doping
profile are determined by our local gates, we can also confine single
holes in an {\em npn} configuration in the same device simply by
inverting the gate voltages, shown in figure 2d. In figure 2e we show
the current as a function of the backgate voltage and the voltage on
the splitgates. In the left of the plot, the leads are doped p-type,
and a positive backgate induces a single electron {\em pnp} quantum
dot. In the right of the plot, the leads are doped n-type and a
negative backgate induces a single hole {\em npn} quantum dot. By
adjusting the splitgate voltages, the {\em pn} junction width, and
thus the tunnel barriers, can be tuned while keeping the electron
number fixed (see figure 2f).

In device D2, we use the gates in our design for a different purpose:
here, we rely on less transparent Schottky contacts as incoming and
outgoing tunneling barriers, and now use the backgate and the two
splitgates as three independent local gates to create a double quantum
dot potential in the nanotube with a tunable interdot coupling. Figure
3 shows the current through the device as a function of the two
splitgate voltages. In the lower left and upper right regions of the
plots, the two splitgates dope the two segments of the nanotube with
carriers of opposite sign, resulting in a {\em pn} double quantum dot
with an interdot barrier formed from a {\em pn} junction. In the upper
left (bottom right) corner, the two splitgates dope both sides of the
nanotube p-type (n-type). In figure 3a, $\mathrm{V_{BG}}$ is set to
ground, which gives a potential in the middle of the nanotube that is
attractive for holes but repulsive for electrons. We consequently
observe single dot behaviour for the first hole and weakly coupled
double dot behaviour for the first electron. In figure 3b, we apply a
positive backgate voltage, $\mathrm{V_{BG}}$ = 250 mV. The potential
in the middle of the nanotube is now repulsive for holes: the first
hole enters a weakly coupled double dot, while electrons fill a mostly
single dot potential.  (At some gate voltages, the presence of the
oxide creates a non uniform potential which results in strongly
coupled double dot instead of purely single dot behaviour. See section
S1 of the Supplementary Information for further discussion.)  By changing
$\mathrm{V_{BG}}$, we can continuously tune the interdot coupling in
the few-electron and few-hole regime from weakly coupled double dot to
single dot behaviour.

 In figure 4, we investigate the tunable inderdot coupling in our
double quantum dot more detail by studying current at the (0,1e)
$\leftrightarrow$ (1e,0) triple point transition of a weakly coupled
double quantum dot. In a weakly coupled double quantum dot, current
can only flow at specific values of the gate voltages, known as triple
points, where the levels in the two dots are aligned, allowing an
electron to tunnel from one dot to the
other\cite{VanderWiel2002Electron}. In figures 4a to c,
$\mathrm{V_{BG}}$ is made more negative, creating a larger barrier for
electron tunnelling between the dots, suppressing the current at the
triple point. However, as we sweep $\mathrm{V_{BG}}$ further, shown in
figures 4d and e, the current increases again, despite creating an
even larger barrier for electron tunnelling.

The explanation of this curious increase of the current is a novel
tunnelling process analogous to the tunnelling paradox in high energy
physics proposed by
Klein\cite{Klein1929Die,Katsnelson2006Chiral,Trauzettel2007Spin}.
Specifically, we will define Klein tunnelling as any enhancement of
the tunnelling of an electron through a barrier due the so-called
negative energy solutions (positron states) that arise in relativistic
quantum mechanics (see Supplementary Information for further
discussion).  In figure 4, the enhancement of the interdot coupling we
observe at large tunnel barrier heights is an example of Klein
tunnelling in a carbon nanotube, where now the valance band of the
nanotube plays the role of the negative energy solutions in
relativistic quantum mechanics. What is unique about the data in
figure 4 is that we have created a direct implementation of Klein's
{\em gedanken} experiment in our double quantum dot device, where we
are able to tune continuously from the normal tunnelling regime to the
Klein tunnelling regime simply by changing the barrier height with a
gate voltage.  We have also observed Klein tunnelling for holes (see
Supplementary Information). In figure 4, what we observe is a kind of
``virtual'' Klein tunnelling, where the electron virtually occupies a
state in the empty valance band in order to tunnel from the left to
the right dot, similar to a cotunnelling
process\cite{DeFranceschi2001Electron}. In addition to our
observations in a double quantum dot, the $npn$ data in figure 2 can
be though of as a type of Klein tunnelling in a different regime,
where the valance band is now occupied with holes, and where Klein
tunnelling occurs by the electron sequentially tunneling across the
two $pn$-junctions. This also emphasizes the close relation between
Klein tunnelling in high energy physics and interband tunnelling
phenomena in semiconductor physics, such as Zener tunnelling in
insulators\cite{Zener1934Theory} and direct interband tunnelling in an
Esaki diode\cite{Esaki1958New}.

Analyzing the current at the (0,1e) $\leftrightarrow$ (1e,0)
transition quantitatively using the result from Stoof and
Nazarov\cite{Stoof1996Timedependent,Fujisawa1998Spontaneous}, we
calculate the tunnel rates $\Gamma_L$ and $\Gamma_R$ of the barriers
to the leads, and the interdot tunnel coupling $t_c$, shown in figure
4h. At these gate voltages, we are in the limit of weak interdot
coupling: $t_c \sim 5$ $\mu$V $<< \Gamma_L$, $\Gamma_R \sim$ 0.6
mV. The interdot coupling, $t_c$, is decreased from an initial value
of 9 $\mu$V to a minimum of 3 $\mu$V as a function of backgate, before
the onset of Klein tunnelling results in an increase up to 9 $\mu$V as
we approach gate voltages where an {\em npn} triple dot is
formed. $\Gamma_L$ and $\Gamma_R$ are found to be independent of the
backgate voltage, indicating that the backgate is not influencing the
Schottky barrier transparency.

Finally, we comment that although we are in the appropriate double
quantum dot coupling regime, we have not found evidence of spin
blockade at any of the expected transitions\cite{Ono2002Current}. (A
parallel magnetic field of 1.5T was applied to ensure that the
nanotube valley degeneracy was lifted). One possible explanation for
this is a singlet-triplet splitting in the (0,2e) state that is much
smaller than the 3 mV single particle spacing we observe in the single
electron quantum dot. This could be an indication of Wigner crystal
formation\cite{Deshpande2008Onedimensional}: in a Wigner crystal, the
electron wavefunction overlap is very small, and consequently the
single-triplet splitting is strongly suppressed. This possibility will
be investigated further using devices with more gates, which could
allow us to probe the Wigner crystallization transition by tuning the
quantum dot confinement potential.

We have presented a new technique for confining single electrons and
single holes in ultra-clean carbon nanotubes. By eliminating disorder
and incorporating local gates, a new level of control over single
electron confinement has been achieved, allowing us to observe a novel
type of tunnelling in a single electron carbon nanotube device. While
our motivation for such a device comes from the spin physics of carbon
nanotubes\cite{churchill-cond-mat}, the fabrication itself could have a much broader use in
carbon nanotube applications, such as electrically doped {\em pn}
junctions for carbon nanotube optical
emission\cite{Misewich2003Electrically}, where low disorder and
multiple gates for electrical control of {\em pn} junctions could
allow the development of new types of optically active devices.

\vspace{2em}

\setlength{\parindent}{0em}

{\bf \large Methods}

Fabrication begins with a p++ Si wafer with 285 nm of thermal silicon
oxide. On top of this, a 50 nm thick n++ polysilicon gate layer is
deposited, followed by a 200 nm LPCVD-TEOS oxide layer. Using
electron-beam lithography and dry etching, a trench of approximately
300 nm deep is etched, forming the two splitgates from the n++
Si gate layer. A 5/25 nm W/Pt layer is deposited to serve as source
and drain contacts, and nanotubes are then grown from patterned Mo/Fe
catalyst\cite{Kong1998Synthesis}. In about half of the devices, a
single carbon nanotube is suspended across the trench making
electrical contact to the source and drain. Transport through the
devices is characterized at room temperature, and selected devices are
cooled to $<$300 mK for low temperature transport measurements. In
total, we have measured 11 devices at low temperatures, of which 4
reached the single electron regime. Here we present data from two
small bandgap devices: D1 with L = 1.5 $\mu$m, W = 300 nm and bandgap
E$_\mathrm{g}$ = 60 mV, and D2 with L = 300 nm, W = 500 nm and
E$_\mathrm{g}$ = 25 mV, where bandgaps are determined by subtracting
the charging energy from the size of the empty dot Coulomb diamond.

\bibliography{paper}{}

\setlength{\parindent}{0pt}

\vspace{2em} 

{\bf \large Acknowledgments}

It is a pleasure to acknowledge P. L. McEuen for the suggestion of
using {\em pn} junctions as tunable barriers, as well as D. Loss,
T. Balder, I. T. Vink, R. N. Schouten, L. M. K. Vandersypen, and
M. H. M. van Weert for useful discussions and suggestions. Supported
by the Dutch Organization for Fundamental Research on Matter (FOM),
the Netherlands Organization for Scientific Research (NWO), and the
Japan Science and Technology Agency International Cooperative Research
Project (JST-ICORP).

\vspace{2em}

{\bf \large Additional Information}

Supplementary information accompanies this paper at
www.nature.com/naturenanotechnology. Reprints and permission
information is available online at
http://npg.nature.com/reprintsandpermissions/.  Correspondence and
requests for materials should be addressed to G.A.S.

\vspace{2em}

The authors declare that they have no competing financial interests.

\pagebreak

\renewcommand{\baselinestretch}{1.3}

\begin{figure}
\begin{center}
\includegraphics{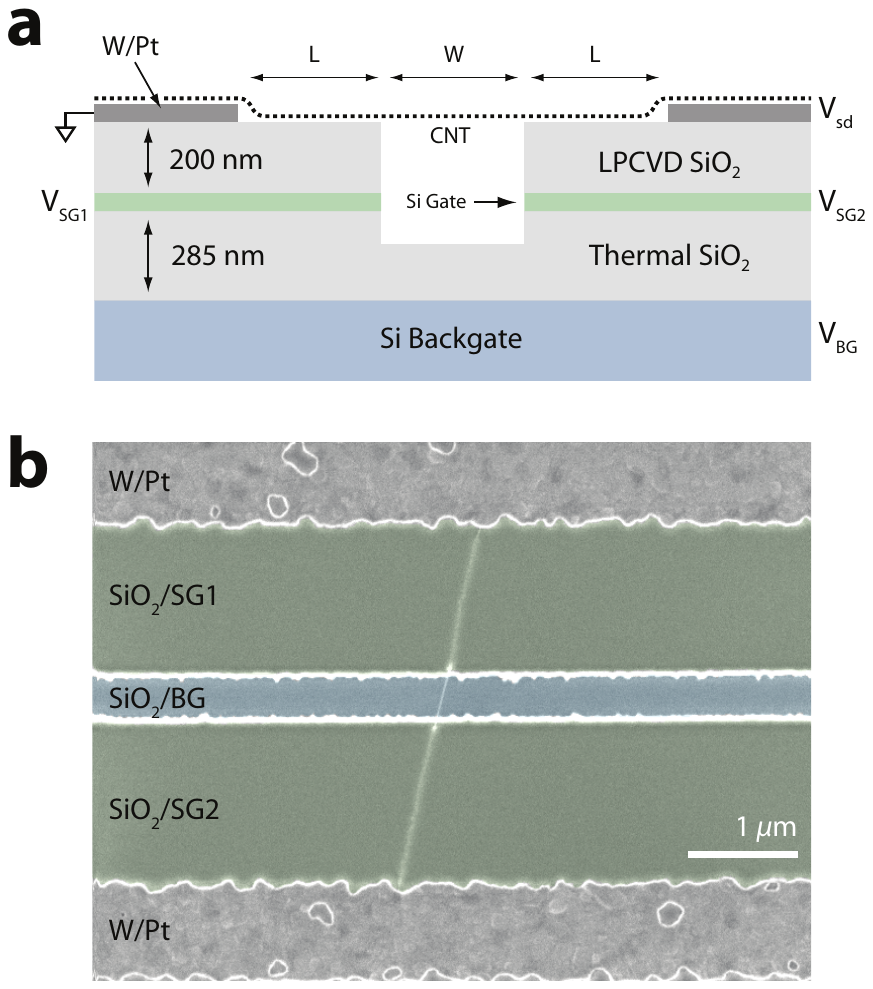}
\end{center}
\caption{ \textbf{Integrating local gates with ultra-clean carbon
nanotubes. a}, A schematic of the device. A predefined trench is
etched to create two splitgates from a 50 nm thick n++ polysilicon
gate layer between two silicon oxide layers. A Pt metal layer is
deposited to act at as source and drain contacts, and a nanotube is
then grown from patterned catalyst. Device D1 has L = 1.5
$\mathrm{\mu}$m, W = 300 nm, and D2 has L = 300 nm, W = 500
nm. \textbf{b}, In a subset of devices, a single nanotube bridges the
trench, contacting the metal source and drain electrodes, as shown in
this colourised SEM micrograph. The micrograph shows an example of a
device with the same dimensions as device D1.}
\end{figure}

\begin{figure}
\begin{center}
\includegraphics{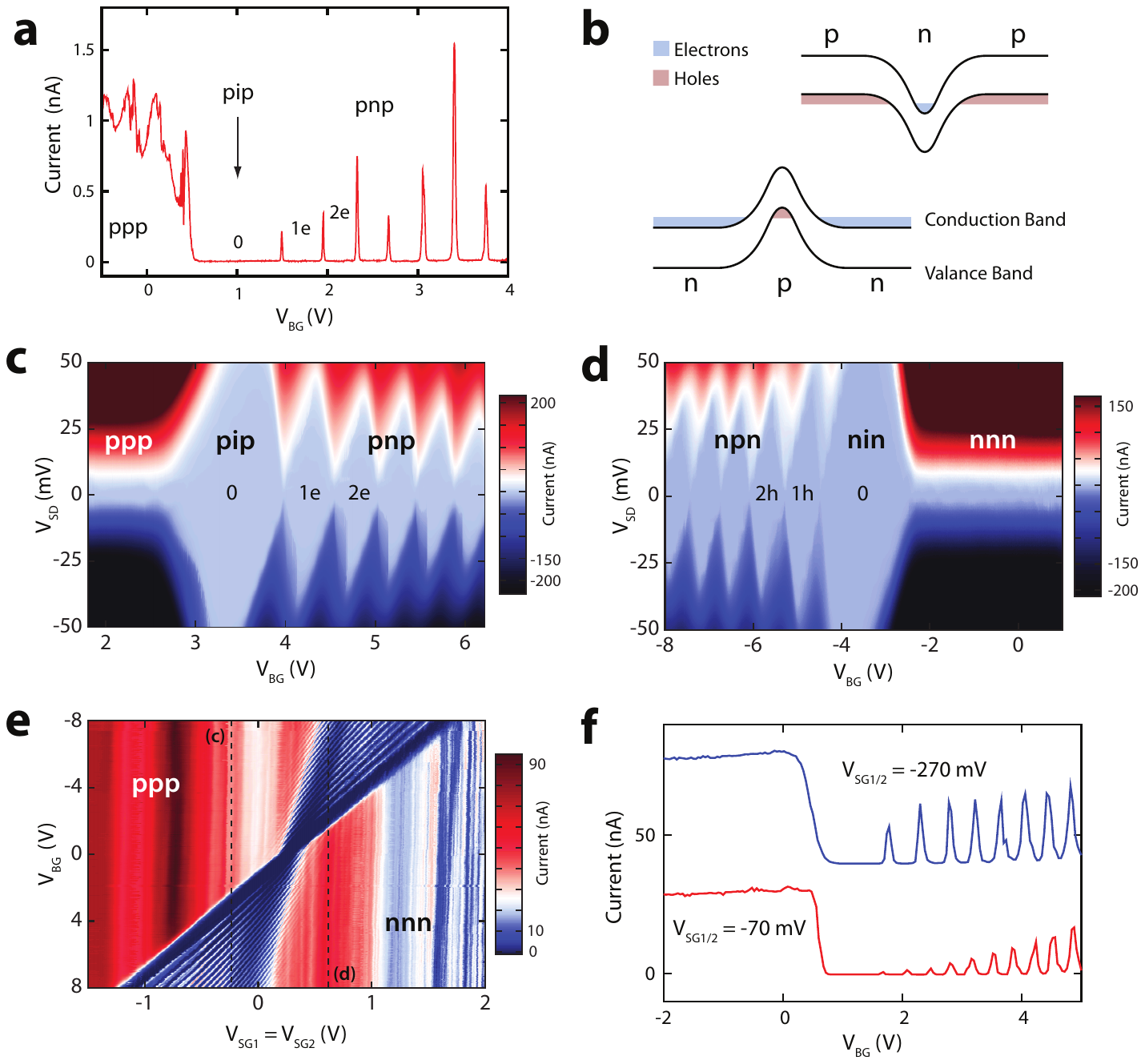}
\end{center}
\caption{ \textbf{Gate defined single-electron and single-hole quantum
dots. a}, Coulomb peaks of a {\em pnp} quantum dot in device D1 taken
at a V$_\mathrm{SG1}$ = V$_\mathrm{SG2}$ = -50 mV and
V$_{\mathrm{sd}}$ = 1 mV.  The splitgates are used to dope the NT
source and drain leads with holes. As V$_{\mathrm{BG}}$ is swept from
negative to positive voltages, the suspended segment is depleted
giving a {\em pip} configuration, followed by a {\em pnp}
configuration as single electrons are filled in an n-type quantum dot,
as illustrated in the energy diagrams in \textbf{b}. \textbf{c},
Stability diagram of the {\em pnp} dot: the charging energy of the
first electron $\mathrm{E_c^{1e} \sim 40}$ meV is remarkably large due
to the weak capacitive coupling of the suspended segment to the gates
and the metal source drain layers.  \textbf{d}, The potential
landscape in the device can be completely controlled by the gate
voltages: by reversing the gate voltages, single holes are confined in
a {\em npn} configuration. \textbf{e}, A 2D plot showing backgate
sweeps at different splitgate voltages and V$_\mathrm{SD}$ = 10
mV. The two splitgates are set to the same voltage. The stability
diagrams in \textbf{c} and \textbf{d} are taken at
V$_{\mathrm{SG1/2}}$ values indicated by the arrows. (Resonances from
residual disorder in the long NT leads can be seen as oscillations as
a function of V$_{\mathrm{SG1/2}}$ in the {\em ppp} and {\em nnn}
configurations.)  \textbf{f}, Using the splitgates, we can tune the
width of the {\em pn} junction depletion region, and hence the tunnel
barriers: at V$_\mathrm{SG1}$ = V $_\mathrm{SG2}$ = -70 mV, the
potential from the splitgates is shallow, giving a wide depletion
region and a current of 0.5 nA for the first electron Coulomb peak at
V$_\mathrm{SD}$ = 10 mV. At V$_\mathrm{SG1}$ = V $_\mathrm{SG2}$ =
-270 mV, the potential across the {\em pn} junction is steeper, now
giving a narrower depletion region and a current of 13 nA for the
first electron. (The V$_{\mathrm{SG1/2}} =$ -270 mV trace has been
offset in V$_\mathrm{BG}$ and in current.) }
\end{figure}

\begin{figure}
\begin{center}
\includegraphics{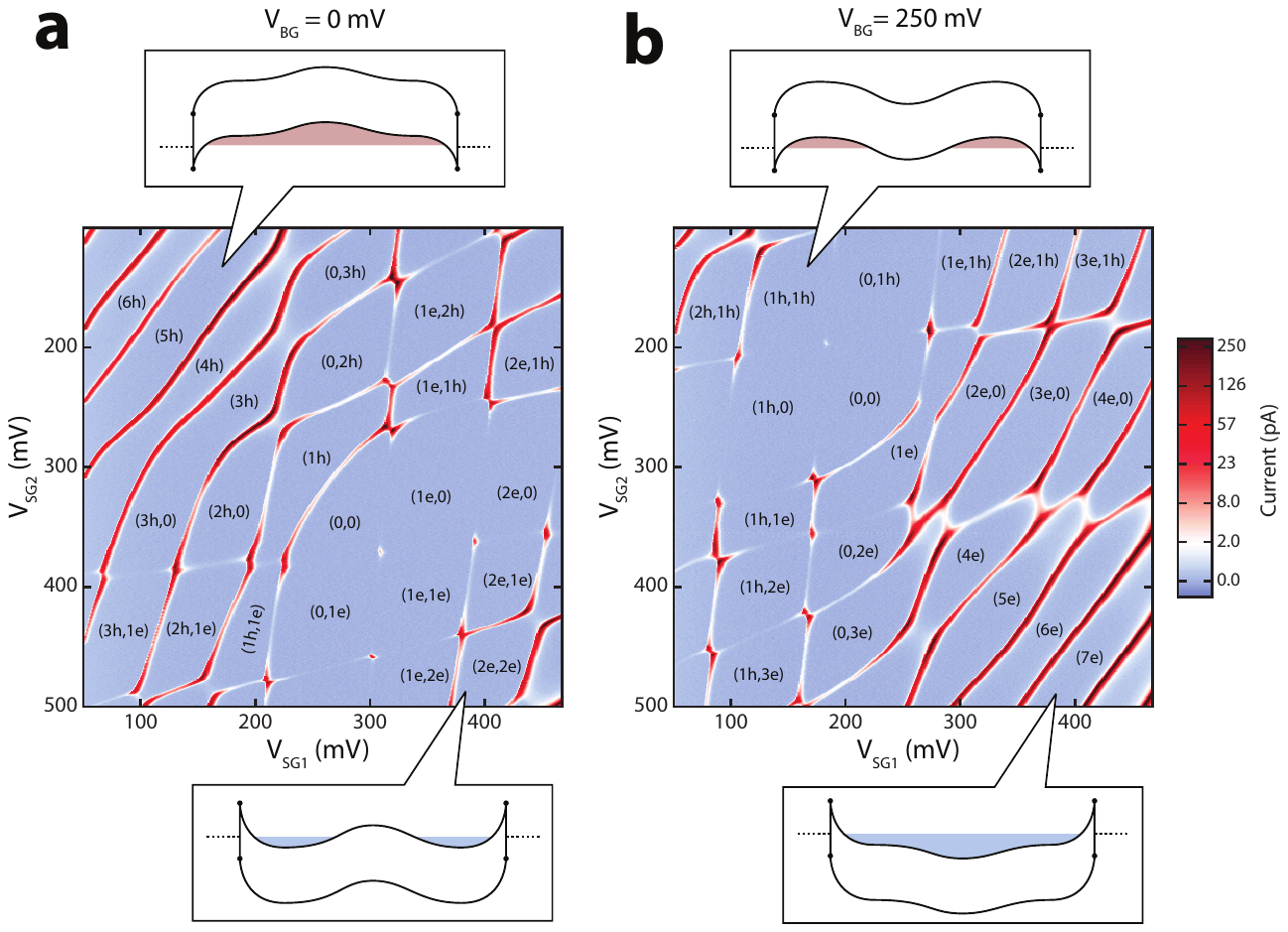}
\end{center}
\caption{ \textbf{A tunable double quantum dot in the few-electron and
few-hole regime.} Current as a function of the two splitgate voltages
at V$_\mathrm{SD}$ = 0.5 mV for device D2. In device D2, electrons are
confined in the nanotube by Schottky barriers at the metal contacts,
with a potential that is tunable using the three gates. Electron and
hole occupation numbers are determined from the transition to a {\em
pn} double quantum dot, as described in the Supplementary
Information. \textbf{a}, V$_{\mathrm{BG}}$ = 0. At this voltage, a
barrier for electrons is induced in the middle of the
device. Electrons are added to a weakly coupled double dot potential,
while holes are added to a single dot potential. \textbf{b},
V$_{\mathrm{BG}}$ = 250 mV. A more positive V$_{\mathrm{BG}}$ creates
a double dot potential for holes and a single dot potential for
electrons. The interdot coupling for both the electron and the hole
double dot can be tuned continuously using the backgate voltage.}
\end{figure}

\begin{figure}
\begin{center}
\includegraphics{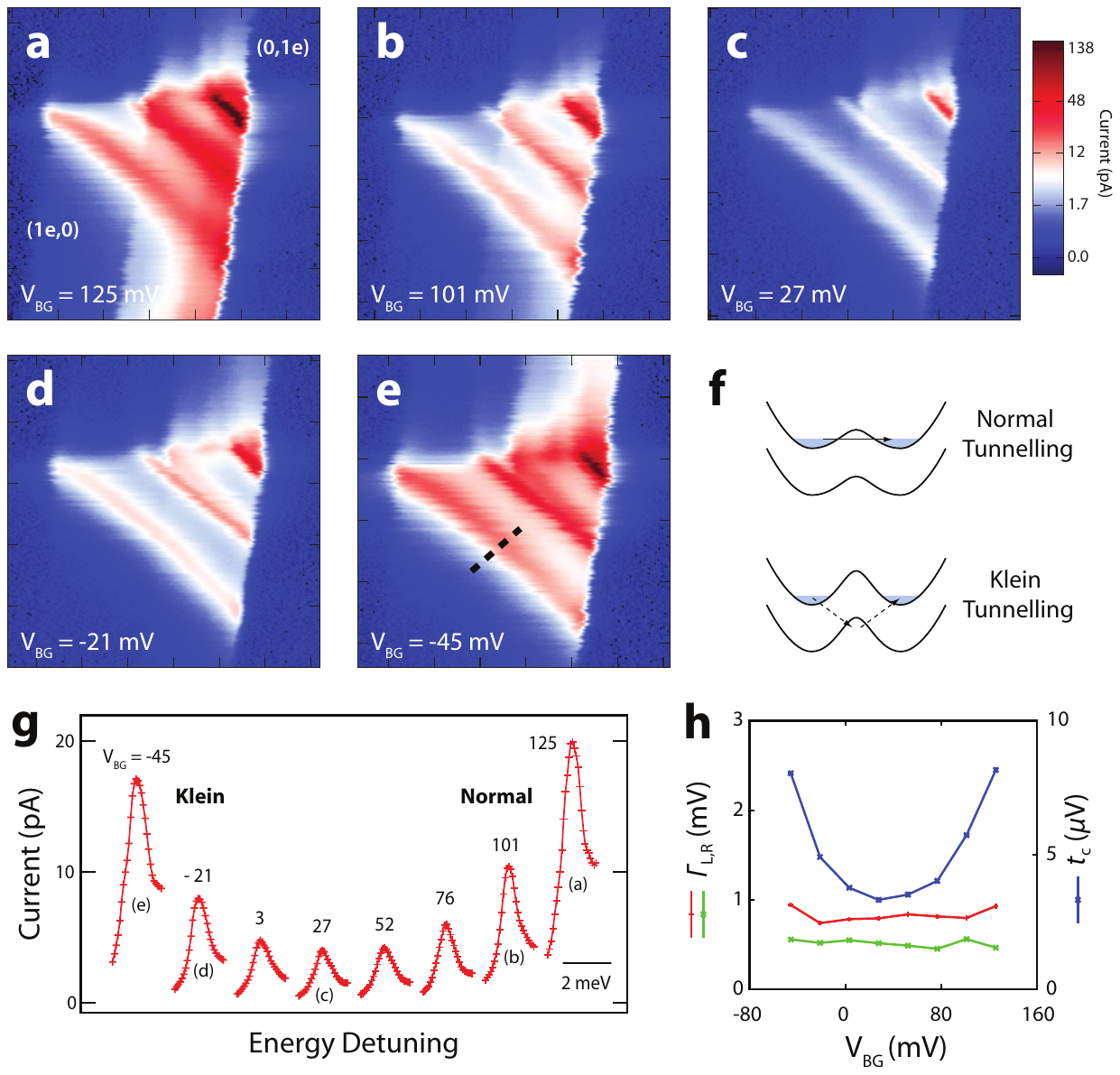}
\end{center}
\caption{ \textbf{Klein tunnelling in a single electron double quantum
dot.} Current at the (1e,0) $\leftrightarrow$ (0,1e) triple point for
a single electron double quantum dot at V$_\mathrm{SD}$ = 5 mV. (Note
that the interdot capacitance $E\mathrm{_c^{inter} \sim 0.2}$ mV is
much smaller than the bias, and thus the triple point bias triangles
for the electron and hole cycle\cite{VanderWiel2002Electron} strongly
overlap.) Transitions to the excited state of the outgoing dot are
visible as lines in the triangle running parallel to the baseline give
a quantized level spacing of 3 mV, consistent with a dot length of
$\sim$500 nm. In \textbf{a} through \textbf{c}, the backgate is made
more negative, creating a larger barrier for electron tunnelling. As a
result, the current through the double dot is decreased. In \textbf{d}
and \textbf{e}, however, the current begins to increase again despite
a larger barrier for electron tunnelling. \textbf{f}, This increase in
current results from tunnelling of an electron below the barrier
through a virtual state in the valence band, analogous to Klein
tunnelling in high energy physics. \textbf{g}, Line cuts of the triple
point data in \textbf{a}-\textbf{e} showing the current for the ground
state baseline transition at different backgate voltages. The line
cuts are taken along the dashed line in \textbf{e}. The x-axis shows
the distance along this line converted into the energy detuning
between the left and right dot ground state levels. For the rightmost
traces, interdot tunnel coupling is mediated by normal electron
tunnelling, while for the leftmost traces, Klein processes provide the
interdot tunnel coupling. \textbf{h}, Parameters from a fit to the
Stoof-Nazarov equation. The interdot tunnel coupling initially
decreases as the barrier height increases (V$_{\mathrm{BG}}$ = 125 to
27 mV), and then increases due to the onset of Klein tunnelling as the
barrier height becomes comparable to the bandgap (V$_{\mathrm{BG}}$ =
27 to -45 mV).}
\end{figure}

\end{document}


\setlength{\parindent}{0em}

{\sf \LARGE Tunable few-electron double quantum dots and Klein tunnelling in
ultra-clean carbon nanotubes: Supplementary Information}

\vspace{0.5em}

G. A. Steele$^1$, G. Gotz$^1$, \& L. P. Kouwenhoven$^1$

\vspace{1em}

{\em $^1$Kavli Institute of NanoScience, Delft University of Technology, PO Box 5046, 2600 GA, Delft, The Netherlands.}

\vspace{1em}

\setlength{\parindent}{2em}

\section*{S1 Determining electron numbers}

Absolute electron numbers in the device are identified by the
transition from a {\em nn} or a {\em pp} single dot to a {\em pn} or
{\em np} double dot, as shown in figure S1. For example, at
V$_\mathrm{SG1} = -250$ mV and V$_\mathrm{SG2} \sim 260$ mV, we remove
the last hole from the right side of the nanotube, (p,p) $\rightarrow$
(p,0). As we sweep V$_\mathrm{SG2}$ further, at V$_\mathrm{SG2} \sim
380$ mV, we fill an electron into the right dot. Here we see an abrupt
transition from single dot behaviour to double dot behaviour,
signaling the transition to a (p,n) double dot. This transition allows
us to clearly identify the electron numbers in the device. The
electron number assignment was also confirmed by large bias Coulomb
diamond measurements such as those shown in figure 2 of the main text.

\begin{figure}
\begin{center}
\includegraphics[width=6in]{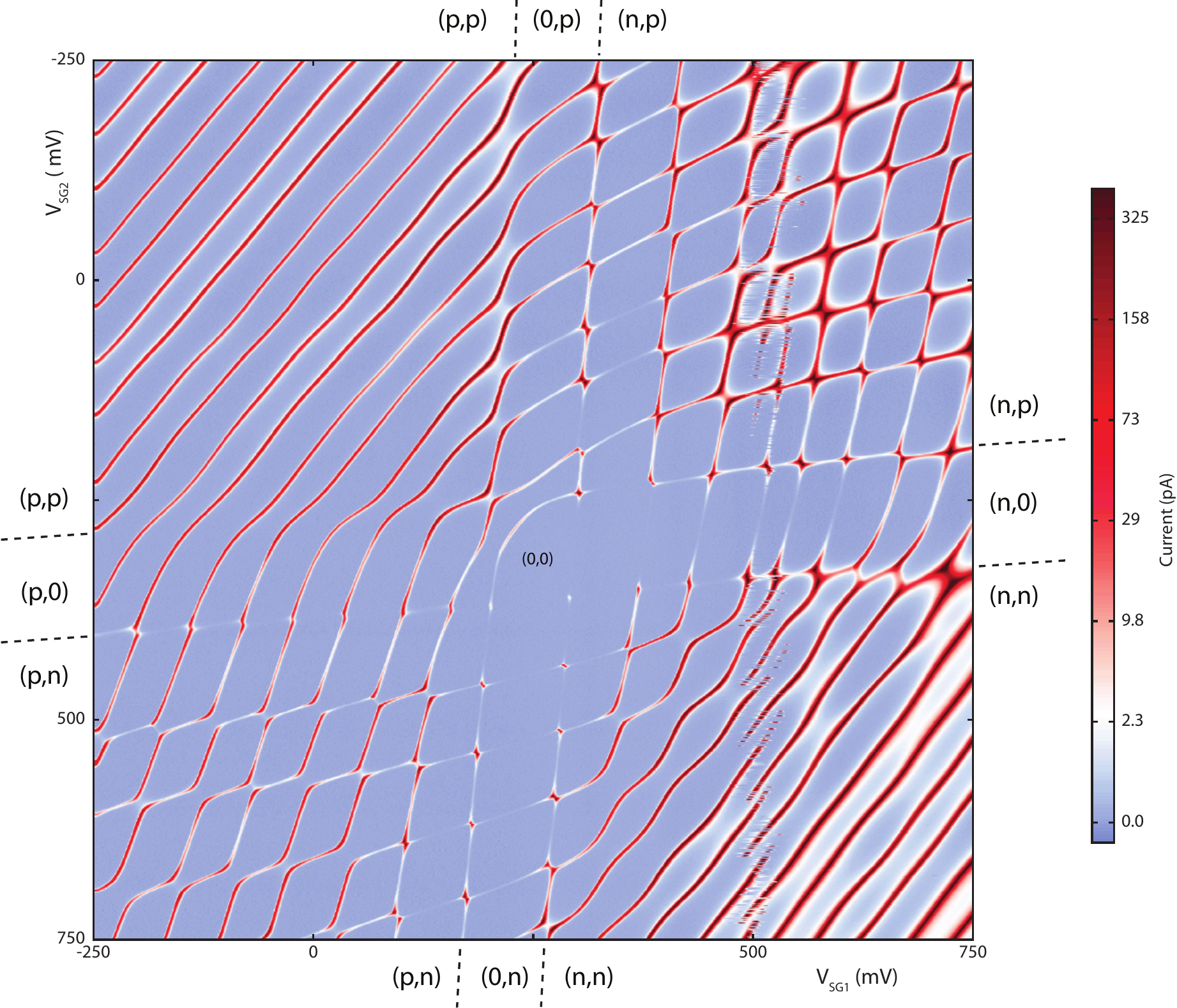}
\end{center}
\caption{A 2D splitgate sweep over a larger range used to determine
electron numbers from the transition to a {\em pn} double quantum
dot. Data is taken at V$_\mathsf{BG}$ = 50 mV and V$_\mathsf{SD}$ =
0.5 mV.}
\end{figure}

At V$_\mathrm{SG1} \sim 500$ mV, the device suffers from a
``switch'' in gate voltage: this switch, which appeared on the third
cooldown of the device, is likely due to a charge trap in the
oxide. Aside from this, the device is extremely stable. It is also
very robust with respect to thermal cycling: after 2 thermal cycles
including exposure to air, the barrier transparencies were unchanged
and the position of the first Coulomb peak moved by less than 50 mV in
gate space.

Note also that although the backgate voltage used in figure S1 should
result in single dot behaviour for holes, the data show some bending of
the Coulomb peak trajectories along the (p,p) to (0,p) and (p,0)
transitions, indicating a strongly tunnel coupled double dot type of
behaviour. It is also visible along the (n,n) to (n,0) and (0,n)
transitions in figure 3(b) of the main text, and at higher electron
numbers in figure S1. This results from a somewhat non uniform
potential induced by the presence of the oxide under part of the tube,
likely due to a combination of trapped charges in the oxide and the
abrupt change in dielectric constant.

\section*{S2 Stoof-Nazarov Equation}

\begin{figure}
\begin{center}
\includegraphics[width=6in]{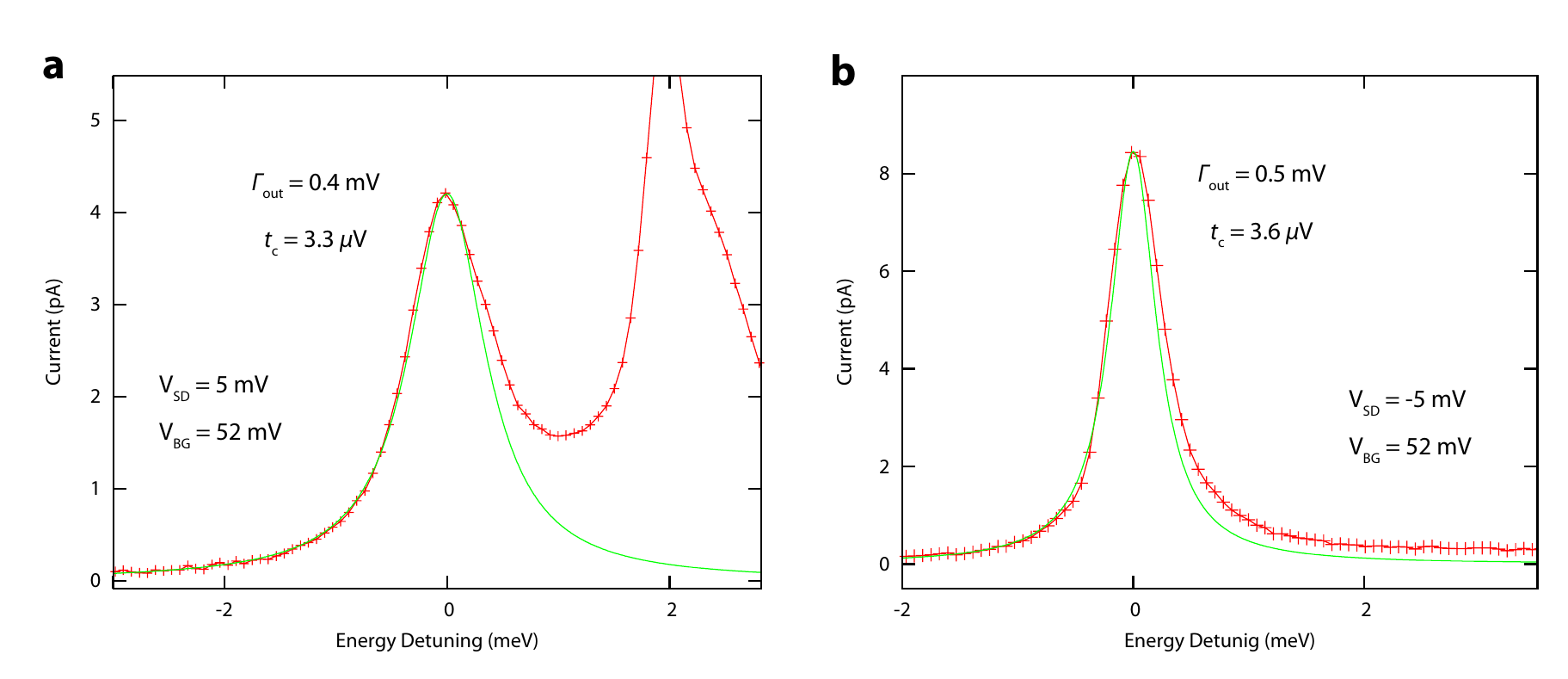}
\end{center}
\caption{Fit of (0,1e) $\leftrightarrow$ (1e,0) transition at $\mathsf{V_{BG}}$ = 52
mV to the Stoof-Nazarov theoretical result for \textbf{a}, positive
and \textbf{b}, negative bias. A detuning independent inelastic
contribution to the current of 350 fA is clearly visible in the
reverse bias trace. This inelastic current is also present in
\textbf{a}, but is more difficult to identify due to a nearby excited
state of the outgoing dot in forward bias.}
\end{figure}

To analyze the data quantitatively, we fit the current at the ground
state to ground state transition along the baseline of the triple
point bias triangle as a function of energy detuning $\epsilon$ to the
expression from Stoof and
Nazarov\cite{Stoof1996Timedependent,Fujisawa2000Inelastic}. By
performing such an analysis, we are able to isolate the contribution
of the middle tunnel barrier from the measurement of the current
through the double quantum dot. For a interdot tunnel coupling $t_c$
and tunnel rates $\Gamma_{L,R}$ to the left and right leads, the
elastic current in a double quantum dot is given by:
\begin{equation}
I_{el}(\epsilon) = \frac{e t_c^2 \Gamma_R}{t_c^2(2 + \Gamma_R/\Gamma_L) + \Gamma_R^2/4 + (\epsilon/h)^2}
\end{equation}
In the limit of weak interdot tunnel coupling, $t_c <<
\Gamma_L,\Gamma_R$, this reduces to a simple Lorentzian line shape of
the form:
\begin{equation}
\label{sn}
I_{el}(\epsilon) = \frac{ 4 e t_c^2 / \Gamma_R}{1 +(2\epsilon/\Gamma_R h)^2}
\end{equation}
A fit of the data to equation \ref{sn} for a single electron double dot is
shown in figure S2. The fit was performed for $\epsilon<0$ to isolate
the purely elastic contribution to the current. For $\epsilon>0$, the
fit deviates from the Lorentzian lineshape due to inelastic
processes\cite{Fujisawa1998Spontaneous}.

\section*{S3 Relativistic tunnelling through a barrier and the Klein
Paradox}

Consider an electron of energy $E$ and momentum $\hbar k$ incident on
a square barrier of height $V$ as shown in figure S3. We are
interested in the probability that the electron is transmitted to
$x>L$ using the Dirac equation. The solutions of the Dirac equation
have two branches\cite{perkins}: a set of positive energy solutions
with $E>0$ and a set of negative energy solutions with $E<0$. The two
branches are separated by an energy gap $2mc^2$. The vacuum state is
interpreted as having the negative energy solutions filled with
electrons (the ``Dirac sea''), and a hole in the Dirac sea is then
interpreted as a positron. For a barrier height that is small compared
to $2mc^2$, shown in figure S3(a), the Dirac equation gives a
wavefunction that decays exponentially inside the barrier: for an
incident energy $E \ll V$, the probability of the electron tunnelling
to the region $x>L$ is small. This is also what is predicted by the
non-relativistic Schroedinger equation.

\begin{figure}
\begin{center}
\includegraphics[width=\textwidth]{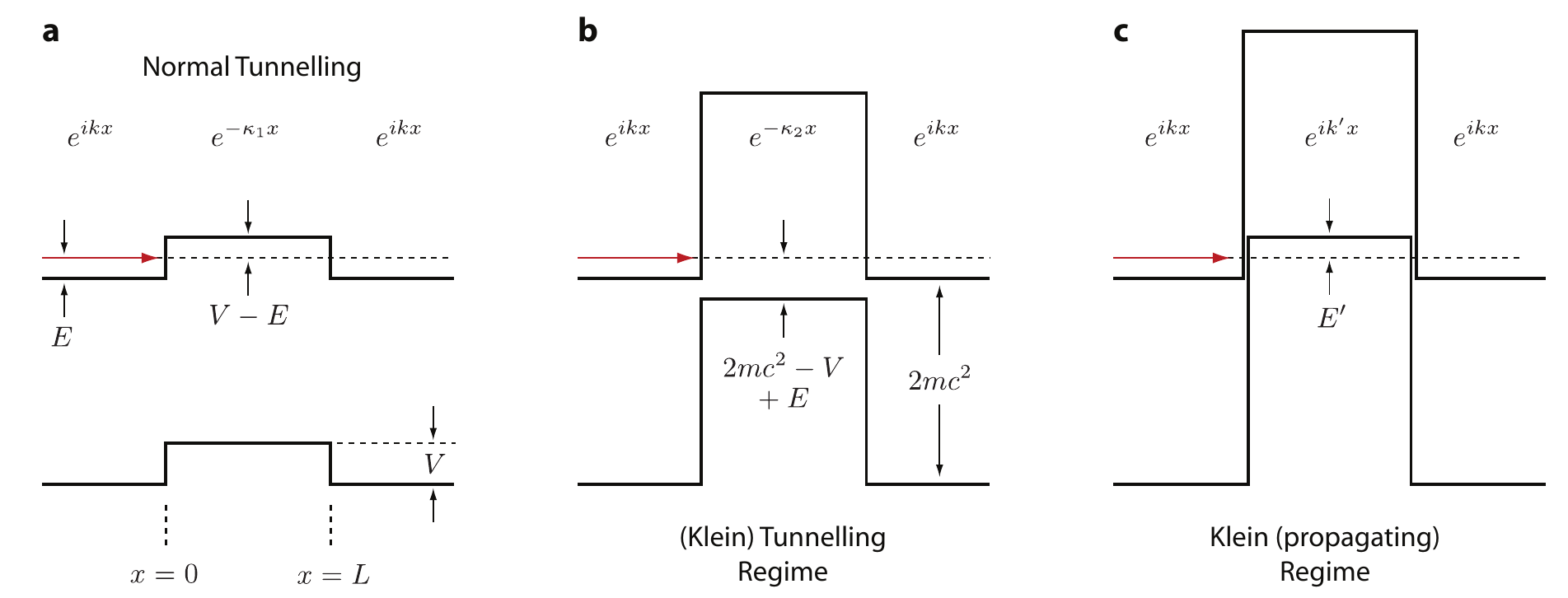}
\end{center}
\caption{Relativistic tunnelling through a barrier. Positive energy
  solutions of the Dirac equation are separated from the negative
  energy solutions by a an energy gap $2mc^2$. We consider the
  probability that an electron incident on a barrier of height $V$ at
  $x=0$ with energy $E$ is transmitted to the region
  $x>L$. \textbf{a,} For $V \ll 2mc^2$, the wavefunction inside the
  barrier decays exponentially with a decay length $\kappa_1 =
  \sqrt{2m(V-E)/\hbar}$, as predicted by the non-relativistic
  Schroedinger equation. We refer to this as the ``Normal'' tunnelling
  regime. \textbf{b,} For $V$ slightly less than $2mc^2$, the
  wavefunction also decays exponentially inside the barrier. However,
  due to the nearby negative energy solutions of the Dirac equation,
  the decay length is now much longer, given by $\kappa_2 =
  \sqrt{2m(2mc^2 - V + E)/\hbar}$, and the transmission probability is
  much higher than that predicted by the Schroedinger equation. We
  refer to this as the (Klein) Tunnelling regime. \textbf{c,} For $V >
  2mc^2$, the electron now propagates inside the barrier without
  decaying by occupying a negative energy solution of the Dirac
  equation. Inside the barrier, the wavefunction is a plane wave
  $e^{ik'x}$ with energy $E' = V - 2mc^2 - E$. We refer to this as the
  Klein (propagating) regime.}
\end{figure}

However, if the barrier height becomes very large, so that $V$ is
comparable to $2mc^2$, the negative energy solutions of the Dirac
equation strongly modify the tunnelling process. In particular, Klein
noticed that for $V \sim 2mc^2$, as shown in S3(b) and S3(c), an
electron moving at non-relativistic speeds incident on the barrier at
position $x=0$ can tunnel to $x>L$ on the other side of the barrier
with nearly unity probability. In the context of the non-relativistic
Schroedinger equation, such a high tunnelling probability would be
completely unexpected, hence the idea of such tunnelling as a paradox.

The tunnelling enhancement can be divided in to two regimes,
illustrated in figures S3(b) and S3(c). We will refer to the first,
illustrated in figure S3(b), as the (Klein) Tunnelling
regime\cite{Bernardini2008Relativistic}. Here, the electron propagates
inside the barrier as an evanescent wave, but the transmission
probability can be very high since the decay length is significantly
longer that that from the Schroedinger equation due to the negative
energy solutions. We will refer to the second regime, shown in figure
S3(c), as the (propagating) Klein regime. Here, the wavefunction in
the barrier is oscillatory in nature and does not decay. Both cases
are examples of what we will call non-classical ``Klein Tunnelling''
in which the electron emerges at $x>L$ with a much higher probability
than that predicted by the Schroedinger equation.

The electronic spectrum of a carbon nanotube at low energies is also
given by a Dirac equation that is the same as that for normal
electrons\cite{Bulaev2008Spinorbit,Trauzettel2007Spin}, but with
$2mc^2$ replaced by the bandgap $Eg$, and the speed of light $c$
replaced by the Fermi velocity of graphene $v_F$
($\sim$0.9$\times$10$^6$ m/s). Free electrons in the Dirac equation
correspond to electrons in the conduction band of the nanotube, and
positrons in the Dirac equation correspond to holes in the valance
band. Thus, it should be possible to observe phenomena analogous to
the two Klein tunnelling regimes of S3(b) and S3(c) in a carbon
nanotube device.

In figure 2 of the main text, we present data demonstrating a single
hole $npn$ and single electron $pnp$ quantum dot. The current that we
observe at the Coulomb peaks can be considered as an example of the
(propagating) Klein regime illustrated in figure S3(c), where the
potential barrier from our gate voltages is larger than the
bandgap. Figure 2(d) from the $pnp$ configuration corresponds to the
(propagating) Klein regime for positrons, and figure 2(c) from the
$npn$ configuration corresponds to the same regime for electrons. 

In figure 4 of the main text, we show an example of the (Klein)
tunnelling regime of figure S3(b). In the data, we observe a
continuous transition from the normal tunnelling regime to that where
the negative energy solutions of the Dirac equation provide an
enhancement of electron tunnelling, as in the original Klein {\em
gedanken} experiment. We also note that the unusual tunnelling process
shown in figure S3(b), where the decay length in the barrier is
increased due to the negative energy states, has recently been
proposed as a mechanism of generating exchange coupling between two
distant quantum dots in graphene
nanoribbons\cite{Trauzettel2007Spin}. Our experiment demonstrates
explicitly such tunnelling in a carbon nanotube.

\section*{S4 Klein tunnelling for a single hole double quantum dot}

In figure S4(a), we show Klein tunnelling for a single hole double
quantum dot. Qualitatively, the process is the same as that for the
single electron double dot. In the single hole double dot, the tunnel
rates to the leads are smaller by a factor of 2-3 compared to the
single electron double dot: in the device, the Fermi level pinning at
the metal contacts is such that electrons see a smaller Schottky
barrier. This can also be seen in figure S1, where the electron single
dot peaks show more current and broadening than those of the hole
single dot.

\begin{figure}
\begin{center}
\includegraphics[width=\textwidth]{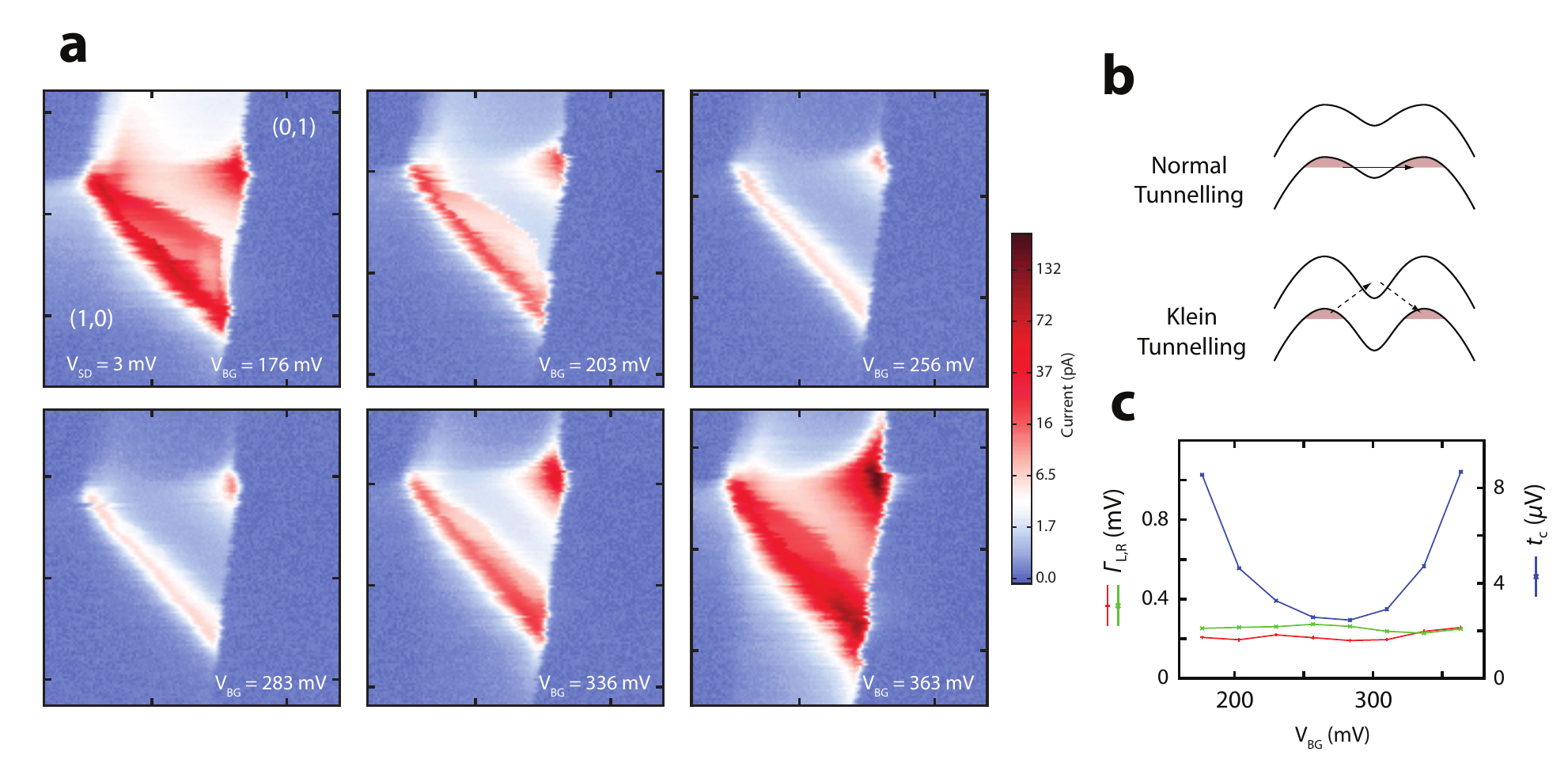}
\end{center}
\caption{\textbf{a,b} Klein tunnelling at the (0,1h) $\leftrightarrow$
(1h,0) single hole double dot transition and \textbf{c}, parameters
from a fit to the Stoof-Nazarov expression.}
\end{figure}

\section*{S5 Extending fabrication to include more gates}

In the device studied in this paper, the incoming and outgoing
barriers of the double dot were formed by Schottky barriers at the
metal contacts. Because we have only three gates, for the double dot
we could not tune the incoming and outgoing barriers independently of
the electron number. As a result, we have significant lifetime
broadening of the energy levels in the double dot configuration.

\begin{figure}
\begin{center}
\includegraphics[width=6in]{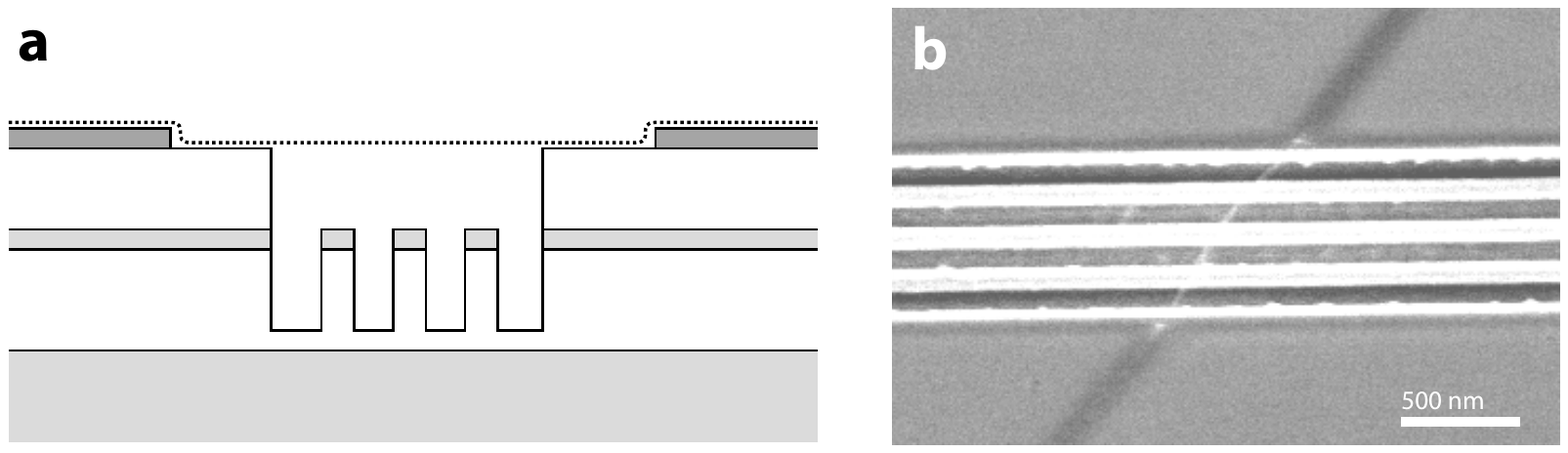}
\end{center}
\caption{\textbf{a}, Schematic of device with more gates in the trench
and \textbf{b}, an SEM micrograph of a test device with a suspended
nanotube.}
\end{figure}
 
To overcome this, we can easily extend the fabrication to include more
gates inside the trench, as shown schematically in figure S5(a).  As
an example of this, a test structure using this type of fabrication is
shown in figure S5(b).

In addition to tuning $\Gamma_{L,R}$ independent of electron number,
these gates could also be used to tune the single dot confinement
potential. If Wigner crystallization is indeed responsible for the
lack of spin blockade in our current device, we should be able to
observe a transition to a finite singlet-triplet splitting by tuning
the confinement energy using these extra gates.

\bibliography{paper}{}


